\documentclass[]{emulateapj}

\slugcomment{ApJ in press}

\shortauthors{Bean et al.}

\begin{document}

\title{The Optical and Near-Infrared Transmission Spectrum of the Super-Earth GJ\,1214b: Further Evidence for a Metal-Rich Atmosphere}

\author{
Jacob L.~Bean\altaffilmark{1,8},
Jean-Michel D{\'e}sert\altaffilmark{1},
Petr Kabath\altaffilmark{2},
Brian Stalder\altaffilmark{1},
Sara Seager\altaffilmark{3},
Eliza Miller-Ricci Kempton\altaffilmark{4,8},
Zachory K.~Berta\altaffilmark{1},
Derek Homeier\altaffilmark{5},
Shane Walsh\altaffilmark{6}, \&
Andreas Seifahrt\altaffilmark{7}
}

\email{jbean@cfa.harvard.edu}

\altaffiltext{1}{Harvard-Smithsonian Center for Astrophysics, 60 Garden Street, Cambridge, MA 02138, USA}
\altaffiltext{2}{European Southern Observatory, Alonso de Cordova 3107, Casilla 19001, Santiago, Chile}
\altaffiltext{3}{Department of Earth, Atmospheric, and Planetary Sciences and Department of Physics, Massachusetts Institute of Technology, Cambridge, MA 02139, USA}
\altaffiltext{4}{Department of Astronomy and Astrophysics, University of California, 1156 High Street, Santa Cruz, CA 95064, USA}
\altaffiltext{5}{Centre de Recherche Astrophysique de Lyon, UMR 5574, CNRS, Universit\'e de Lyon, \'Ecole Normale Sup\'erieure de Lyon, 46 All\'ee d'Italie, F-69364 Lyon Cedex 07, France}
\altaffiltext{6}{Magellan Fellow, Australian Astronomical Observatory and Curtin Institute of Radio Astronomy, Curtin University, GPO Box U1987, Perth, WA 6845, Australia.}
\altaffiltext{7}{Department of Physics, University of California, One Shields Avenue, Davis, CA 95616, USA}
\altaffiltext{8}{Sagan Fellow}

\begin{abstract}
We present an investigation of the transmission spectrum of the 6.5\,$M_{\oplus}$ planet GJ\,1214b based on new ground-based observations of transits of the planet in the optical and near-infrared, and on previously published data. Observations with the VLT\,+\,FORS and Magellan\,+\,MMIRS using the technique of multi-object spectroscopy with wide slits yielded new measurements of the planet's transmission spectrum from 0.61 to 0.85\,$\mu$m, and in the $J$, $H$, and $K$ atmospheric windows. We also present a new measurement based on narrow-band photometry centered at 2.09\,$\mu$m with the VLT\,+\,HAWKI. We combined these data with results from a re-analysis of previously published FORS data from 0.78 to 1.00\,$\mu$m using an improved data reduction algorithm, and previously reported values based on \textit{Spitzer} data at 3.6 and 4.5\,$\mu$m. All of the data are consistent with a featureless transmission spectrum for the planet. Our $K$-band data are inconsistent with the detection of spectral features at these wavelengths reported by Croll and collaborators at the level of 4.1\,$\sigma$. The planet's atmosphere must either have at least 70\% H$_{2}$O by mass or optically thick high-altitude clouds or haze to be consistent with the data.
\end{abstract}

 
\section{INTRODUCTION}
Recent results from the \textit{Kepler} mission indicate that exoplanets intermediate in size between the terrestrial and the ice giant planets in the Solar System are common at small orbital separations from main sequence stars \citep{borucki11,howard11}. However, despite the wealth of knowledge of the sizes and masses for these planets that will be obtained from the \textit{Kepler} data and follow-up efforts, there will still be a substantial gap in our understanding of these objects. This is because knowledge of only the mass and radius of planets in this regime is not sufficient to determine their bulk composition due to degeneracies that exist in theoretical models \citep{adams08,rogers10a}. Specifically, most planets with masses and radii in the intermediate-size range can be modeled equally well by substantially different combinations of mass and composition for the interior and surrounding atmosphere.

Fortunately, there is a way to break the degeneracies between different models for a planet when a unique solution for the bulk composition can not be found. The method is to obtain constraints on the planet's atmospheric composition \citep{millerricci09,millerricci10}. This information provides a boundary condition to further constrain both rigorous physical models and also cosmogonical arguments about likely origin and evolutionary scenarios. Therefore, detailed characterization of representative objects to further constrain their bulk compositions is an important complement to the statistical study of the occurrence rate of populations that is being enabled by \textit{Kepler}. 

The transiting planet GJ\,1214b \citep{charbonneau09} is one object that is potentially a key to unlocking the mystery of the bulk compositions of intermediate-size planets using atmospheric studies. Its determined mass (6.5\,M$_{\oplus}$) and radius (2.65\,R$_{\oplus}$) put it firmly in the degenerate region of parameter space for interior and atmosphere mass and compositions. The only thing that can be said with a high degree of confidence based on just the mass and radius of the planet is that it must have a substantial atmosphere because it is too large to be composed solely of solid material \citep{rogers10b,nettelmann11}. 

\citet{rogers10b} identified three possible origins for the gas layer on GJ\,1214b and they presented distinct interior structure models based on these scenarios that are consistent with its measured mass and radius. The model scenarios are (1) a ``Mini-Neptune'' planet that is composed mainly of solid rock and ice and that has a significant primordial atmosphere accreted from the protoplanetary nebula and with a composition roughly similar to that of the Sun (i.e., hydrogen-dominated), (2) a ``Water World'' planet composed primarily of water ice and having a secondary gas envelope formed by sublimation and dominated by water vapor, and (3) a true ``Super-Earth'' planet composed of purely rocky material with a secondary atmosphere formed by outgassing and probably composed mainly of hydrogen. Distinguishing between these competing models for GJ\,1214b would have important consequences because they each imply a very different formation and evolutionary history for the planet. Constraining the formation and evolutionary history for this planet would be an important step in our quest to obtain a general understanding of the intermediate-size planets for which this object appears to be an archetype.

The most interesting characteristic of GJ\,1214b is that it orbits a very small M dwarf, and thus the system has a planet-to-star radius ratio comparable to a Jupiter-size planet transiting a Sun-like star. This quality makes it the most feasible known intermediate-size planet for atmospheric studies using transit spectroscopy techniques. The three models proposed for the planet would also exhibit very different transmission spectrum features. Mini-Neptune and true Super-Earth planets with their hydrogen-dominated, and thus large scale height, atmospheres would exhibit relatively large spectral features in transmission owing to absorption by trace gases like water and methane, and scattering by molecular hydrogen and cloud or haze particles. On the other hand, a Water World planet with a primarily water vapor atmosphere, and thus small scale height, would exhibit relatively small spectral features in transmission. 

We recently obtained measurements of GJ\,1214b's transmission spectrum in the red optical (0.78 -- 1.00\,$\mu$m) with the FORS instrument on the VLT using a new ground-based technique to constrain the planet's atmospheric composition \citep{bean10}. This remarkably featureless spectrum is consistent with a model for a metal-rich atmosphere with a small scale height, and inconsistent with a model for cloud-free hydrogen-dominated atmospheres, which would have a large scale height. Clouds or haze in a hydrogen-dominated atmosphere could conceivably also yield a flat spectrum consistent with the observations. Subsequently, we presented broadband photometric measurements obtained at 3.6 and 4.5\,$\mu$m with \textit{Spitzer} that provided additional constraints on the composition of the atmosphere \citep{desert11}. These data were also consistent with a featureless transmission spectrum when analyzed in isolation and in combination with the FORS data, but the overall interpretation of the flat spectrum remained uncertain due to the possibility of clouds or haze providing a gray opacity source, and non-equilibrium abundances of methane, which is the primary expected opacity source at 3.6\,$\mu$m. 

In contrast to the results showing a featureless spectrum, \citet{croll11a} presented ground-based near-infrared photometry observations that indicated GJ\,1214b's transmission spectrum has a large feature in the $K_{s}$-band (bandpass approximately 2.0 -- 2.3\,$\mu$m) based on the measurement of a substantially deeper transit in that band as compared to the observations at other wavelengths. The main interpretation of the observations in this case is unambiguous; the large scale height of a hydrogen-dominated atmosphere is required for such a feature to be observed. 

\citet{crossfield11} presented high-resolution spectroscopy of GJ\,1214b between 2.1 -- 2.4\,$\mu$m. They detected no spectral features, but could not comment on the absolute depth of the transit at these wavelengths due to the purely differential nature of their data. \citet{crossfield11} came to similar conclusions as \citet{desert11} did based on the \textit{Spitzer} data because methane is also the expected source of wavelength-dependent opacity in the their observational window. 

The observational results showing a featureless transmission spectrum for GJ\,1214b and the result from \citet{croll11a} indicating large spectral features in the near-infrared are not strictly in conflict because they either were obtained at different wavelengths [i.e., the \citet{bean10} and \citet{desert11} studies] or have different sensitivities [i.e., the \citet{crossfield11} investigation]. Also, there is a plausible qualitative explanation for all the observations: a hyrodgen-dominated atmosphere depleted in methane and with clouds or haze opaque at optical wavelengths. However, there is some tension because, for all the plausibility of the qualitative model, it is not supported by physical calculations at this point, and it has to be finely tuned to match the observations.

Given the outstanding issues in our understanding of GJ\,1214b's atmosphere, we were motivated to conduct further observational studies of its transmission spectrum to improve our understanding of this important world. In particular, we aimed to make independent measurements of the planet's transmission spectrum in the $K$-band region using a different technique than \citet{croll11a} used, and also to make measurements further in the blue part of the optical to search for the signature of scattering from cloud, haze, or gas particles. We present here the results of these investigations. In \S2 we describe our observations and data reduction. In \S3 we describe our analysis of the data. We compare our new measurements of the planet's transmission spectrum to previous measurements and theoretical models in \S4. We conclude in \S5 with a summary of the results.

\section{Data}
\subsection{Magellan\,+\,MMIRS}
\subsubsection{Observations}
We observed transits of GJ\,1214b on 2011 May 15 and 18 using the MMIRS instrument \citep{mcleod04} on the Magellan (Clay) telescope at Las Campanas Observatory. We used the same multi-object spectroscopy approach for the MMIRS observations as was used for the FORS observations of GJ\,1214b that we presented previously \citep{bean10}. We gathered time-series spectra for GJ\,1214b and three other reference stars of similar brightness within 6\arcmin\ using slits with lengths of 30\arcsec\ and widths of 12\arcsec. The off-axis guide and wavefront sensor of the instrument, which is sensitive to wavelengths between 0.6 and 0.9\,$\mu$m, was used to feed corrections to the telescope control and active optics systems. We used an HK grism as the dispersive element and an HK filter to isolate the first order spectra. Complete spectra from 1.23 to 2.48\,$\mu$m with a dispersion of 6.6\,\AA\,pixel$^{-1}$ were obtained for all the objects.

\begin{figure}
\resizebox{\hsize}{!}{\includegraphics{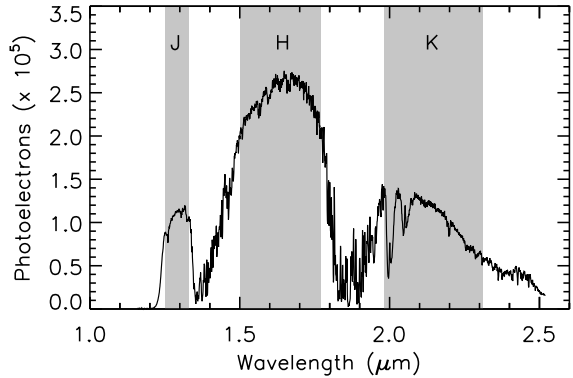}}
\caption{An example spectrum extracted from the MMIRS data for GJ\,1214. The broad bandpasses in the $J$, $H$, and $K$ atmospheric windowns are indicated by the grey boxes.}
\label{fig:mmirs_star_spectrum}
\end{figure}

As described below, only the data obtained on the second night are reliable. The observations for this transit spanned 2.63\,h from UT 06:15 to 08:53. Exposure times were 25\,s, and the overhead per exposure including the read and reset time of the detector was 14\,s. A total of 239 exposures were obtained, 79 of which were in transit. The observations began when the field was at an airmass of 1.20, and by the end the field had set to an airmass of 1.58. Conditions were clear and the seeing varied between 0.5 and 0.9\arcsec\ as estimated by the width of the spatial profiles of the obtained spectra.

\subsubsection{Data reduction}
The MMIRS data were recorded using the ``up-the-ramp'' sampling mode of the detector in which non-destructive reads were obtained every 5\,s and stored in a data cube. In the first step of the data reduction, we subtracted the first of these reads from the others on a pixel-by-pixel basis in each frame to remove the bias and persistence from previous exposures. The estimated dark current of the detector is 0.06\,$e^{-}$\,s\,$^{-1}$\,pixel$^{-1}$, which is negligible for this experiment. After subtracting the first read in the data cube for an exposure, we then fit a slope to the up-the-ramp samples for each pixel to determine the count rate. We arrived at the final estimate of the total pixel counts by multiplying the count rate by the total exposure time. 

The reduction of the MMIRS data cubes gave reasonable results for the second data set, but we noticed a repeating pattern of discontinuities in the residuals when fitting the up-the-ramp samples for the first data set. The origin of this effect is likely a problem in the detector electronics. We chose not to include the first data set in the final analysis because of this problem; in what follows we focus only on the analysis of the data set obtained on the 18th.

After the data processing to collapse the data cubes, we applied spectroscopic flat-field corrections, subtracted the background, and extracted one-dimensional spectra for the objects using the same algorithms as for our previous multi-object spectroscopy observations \citep{bean10}. We established the wavelength scale of the extracted stellar spectra based on spectra of an Ar emission lamp obtained following the transit observation using a copy of the science mask with the slit widths set to 0.5\arcsec. We adopted a third-order polynomial form for the wavelength solution. The wavelength calibration was done independently for each of the four slits, and the maximum dispersion from the fits was 0.3\,\AA. After applying the wavelength solution to the data, we noticed offsets between the positions of telluric absorption lines in the spectra extracted from the first image. These offsets are consistent with a misalignment in the rotation angle between the science mask and the calibration mask. We determined and applied offset corrections to the zero point of the wavelength scale of each object. The maximal correction size was 38\,\AA, which is equivalent to approximately six pixels.

The positions of the spectra on the detector varied in both the spatial and spectral directions during the time-series due to imperfect guiding and instrument flexure, and this led to apparent shifts in the extracted 1d spectra relative to each other. We cross-correlated the spectra for each object relative to the first spectrum in the time-series, and then shifted them by this amount to correct for the effect of the drift. The position of the spectra drifted by 2.5 pixels over the course of the first 25 exposures, and then there was a sudden shift of approximately 3 pixels. The position of data after the large jump remained relatively stable with a rms of 0.15 pixels and only a small long-term drift totaling 0.13 pixels. We removed the first 30 points from the time-series to avoid problems caused by the large instability at the beginning. There is still sufficient out-of-transit baseline (45\,minutes) with these points removed, and the model fits are substantially better for the trimmed light curves.

\begin{figure}
\resizebox{\hsize}{!}{\includegraphics{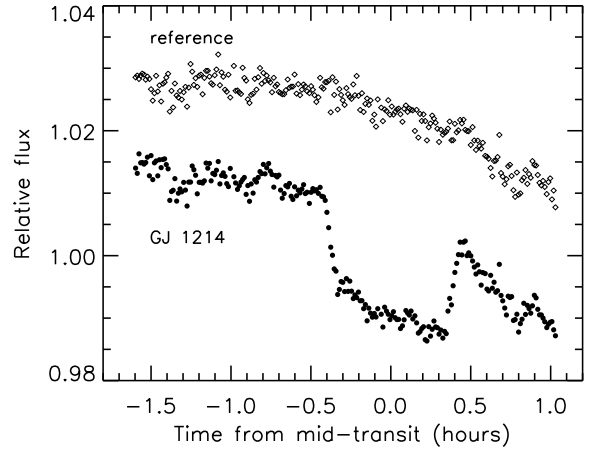}}
\caption{The MMIRS $K$-band raw time-series data for GJ\,1214 (filled circles) and the composite reference (open diamonds).}
\label{fig:mmirs_reference}
\end{figure}

\begin{figure*}
\resizebox{\hsize}{!}{\includegraphics{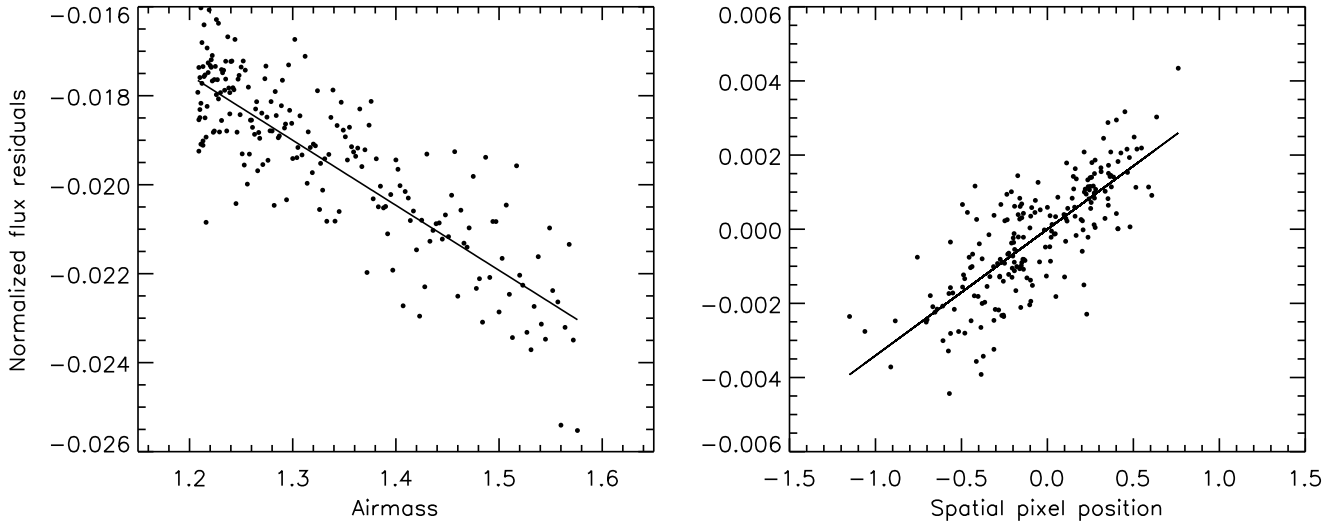}}
\caption{Normalized residuals (circles) from the MMIRS broadband $K$-band light curve fits without the airmass (left panel) and the spatial spectrum position (right panel) decorrelations as a function of those parameters. The decorrelation fits are indicated by the solid lines.}
\label{fig:correlation}
\end{figure*}

An example spectrum of GJ\,1214 is shown in Figure~\ref{fig:mmirs_star_spectrum}. The signal-to-noise ratio (SNR) in the GJ\,1214 spectra estimated from photon counting statistics reaches upwards of 475\,pixel$^{-1}$ in the $H$-band atmospheric window where the maximum number of counts were recorded. We estimate that the resolution of the stellar spectra is approximately 50\,\AA\ (7.5 pixels). This estimate is based on the intrinsic resolving power of the grism (R $\equiv$ $\lambda$/$\Delta \lambda$ $\approx$ 560 for a 1.0\arcsec\ slit), the atmospheric seeing, errors in the correction for the motion of the stars on the slits during the observations, and the uncertainty in the absolute wavelength scale.

\subsubsection{Extraction of the photometric time-series}
We created broadband photometric time-series from the extracted spectra in the $J$, $H$, and $K$ atmospheric windows, and four spectroscopic time-series in the $K$-band of width 0.1\,$\mu$m by summing the obtained spectra within a bandpass. As discussed in \S3, both the broadband and spectroscopic data provide insight to the composition of the planet's atmosphere. We do not feel that the $J$- and $H$-band data are of sufficient quality to justify subdividing them into spectroscopic channels. 

The definitions of the utilized bands and the determined planet-to-star radius ratios ($R_{p}/R_{\star}$) from the light curve modeling (see \S3) are given in Table~\ref{tab:mmirs_band}. The limits of the broad bandpasses were chosen to match standard near-infrared filter curves\footnote[1]{See \url{http://www.cfht.hawaii.edu/Instruments/Filters/wircam.html} for the properties of the filters used by \citet{croll11a}.} as closely as possible to avoid regions with significant telluric water absorption and to enable quick comparison with other results, in particular the $K_{s}$-band measurement of \citet{croll11a}. However, our data do not include the entire range of the standard $J$-band, and the combination of the grism and order-blocking filter we used has substantially different transmission properties than broadband filters. The exact locations of the bandpass edges were also set so that the spectra were not cut in the middle of strong absorption lines. Therefore, appropriately weighted integration of the atmospheric models for the planet are necessary for robust comparison of results even at similar wavelengths.

\begin{deluxetable}{lcc}
\tabletypesize{\scriptsize}
\tablecolumns{3}
\tablewidth{0pc}
\tablecaption{Photometric Bandpasses \& Transit Depths for the MMIRS Data}
\tablehead{
 \colhead{Band} &
 \colhead{Wavelength ($\mu$m)} &
 \colhead{$R_{p}/R_{\star}$\tablenotemark{a}}
}
\startdata
Broadband analysis \\[0.5mm]
$J$ & 1.25 -- 1.33 & 0.1158\,$\pm$\,0.0024 \\
$H$ & 1.50 -- 1.77 & 0.1146\,$\pm$\,0.0014 \\
$K$ & 1.98 -- 2.31 & 0.1158\,$\pm$\,0.0006 \\[2mm]
\hline\\
Spectroscopic analysis \\[0.5mm]
channel 1 & 1.98 -- 2.08 & 0.1156\,$\pm$\,0.0007 \\
channel 2 & 2.08 -- 2.18 & 0.1163\,$\pm$\,0.0007 \\
channel 3 & 2.18 -- 2.28 & 0.1158\,$\pm$\,0.0006 \\
channel 4 & 2.28 -- 2.38 & 0.1163\,$\pm$\,0.0011
\enddata
\tablenotetext{a}{From an analysis with the transit parameters $a/R{\star}$ and $i$ fixed to 14.97 and 88.94\degr, respectively.}
\label{tab:mmirs_band}
\end{deluxetable}

The time-series data for GJ\,1214 in each bandpass were corrected by dividing out the sum of the data for the reference stars. In the subsequent light curve modeling, we found that including all of the reference stars in this step yielded a worse correction than when one of them was excluded. A possible explanation for this is that the spectra for the problematic reference star fell along the boundary between different read-out channels in the detector. The motion of the spectra during the time-series resulted in the spectra for this object moving back and forth across the boundary and this could have introduced additional noise in the data. For the final analysis, we excluded the data for the bad reference star, and we used only the data for the other two reference stars to correct the GJ\,1214 time-series. The estimated photon-limited uncertainties in the photometry for the reference star were propagated through to the final uncertainties in the GJ\,1214 data. An illustration of the reference star correction to the broadband $K$-band data is shown in Figure~\ref{fig:mmirs_reference}.

After correcting the time-series for GJ\,1214 using the reference star data, we identified additional systematics correlated with the airmass of the observations and the positions of the spectra in the spatial dimension on the detector. We fit for linear decorrelations against these two parameters simultaneously with the light curve modeling described in \S3 to remove these effects (two free parameters in addition to the normalization). Adding higher-order terms and decorrelating against airmass and spatial pixel position for the individual stellar fluxes (i.e., GJ\,1214 and the reference stars) did not improve the results. The correlations between the reference star-corrected light curve residuals and the resulting linear fits for these two parameters are shown in Figure~\ref{fig:correlation}.

The MMIRS broadband light curves for GJ\,1214 after all corrections and decorrelations exhibit residuals from a best-fit model with rms of 1532, 887, and 706\,ppm in the $J$-, $H$-, and $K$-bands, respectively. These values are factors of 3.9, 6.0, and 3.4 higher than expected from the photon-limited uncertainties. The $H$-band data exhibit an especially larger-than-expected scatter of unknown origin. We experimented with changing the bandpasses to investigate this issue. Shifting the bandpasses and also using smaller and larger windows did not significantly affect the results. We speculate that perhaps the transparency of the Earth's atmosphere varies on different spatial scales among the near-infrared atmospheric windows; and that this scale, along with the number, color, and spatial distribution of reference stars, sets the fundamental limit when working in the $\lesssim$\,1000\,ppm photon-limited regime in the near-infrared. When binned over 60\,s, the rms of the $K$-band residuals is 443\,ppm. These are the highest-precision ground-based near-infrared transit light curve data that we are aware of. The rms of the residuals for the $K$-band spectroscopic channels range from 923\,ppm to 1552\,ppm, and the scaling from the expected levels of noise range from 2.3 to 2.8.

\begin{figure}
\resizebox{\hsize}{!}{\includegraphics{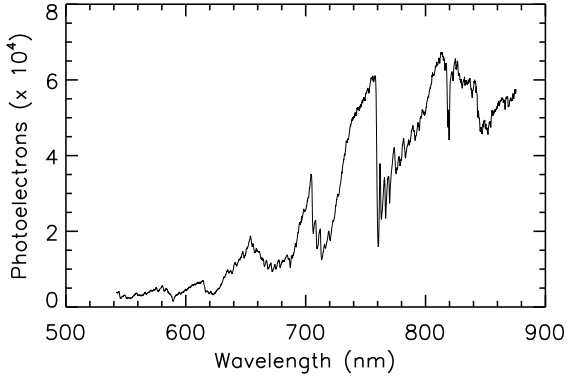}}
\caption{An example spectrum extracted from the FORS blue data for GJ\,1214}
\label{fig:fors_star_spectrum}
\end{figure}

\subsection{VLT\,+\,FORS}
We observed one transit of GJ\,1214b on 2011 July 3 using the multi-object spectroscopy mode of the FORS instrument \citep{appenzeller98} on UT1 of the VLT facility at Paranal Observatory. The instrument setup and data reduction were very similar to the previous FORS observations described in \citet{bean10} and \citet{berta11}, and we limit our discussion here to the unique aspects of the current study. 

We used the 600RI grism with the GG435 filter to obtain spectra over the range 610 -- 850\,nm for GJ\,1214 and five reference stars. Observations began at UT 02:04 and continued for 3.12\,h to UT 05:11. The field was at an airmass of 1.20 at the beginning, rose to an airmass of 1.15, and then set to an airmass of 1.33 by the end. The exposure times were 18\,s, which gave a total duty cycle of 55\,s including 37\,s of overhead. A total of 194 exposures were obtained, 57 of which were during the transit.

Our approach for the data reduction, spectral extraction, and wavelength calibration were the same as for the previous FORS observations \citep{bean10}, with one exception. During the wavelength calibration step, we identified a few lamp lines in the FORS atlas that must have inaccurate assigned wavelengths because they were significantly discrepant from a low-order polynomial dispersion model. When we did not include these lines in the solution, it became clear that a third-order polynomial model for the dispersion was justified instead of the linear model we had used previously and that is also advocated in the instrument manual. The residuals from the third-order dispersion fit with the bad lines ignored are typically 0.005\,nm, whereas the residuals are typically 0.5\,nm from a linear model with the bad lines included.

\begin{figure}
\resizebox{\hsize}{!}{\includegraphics{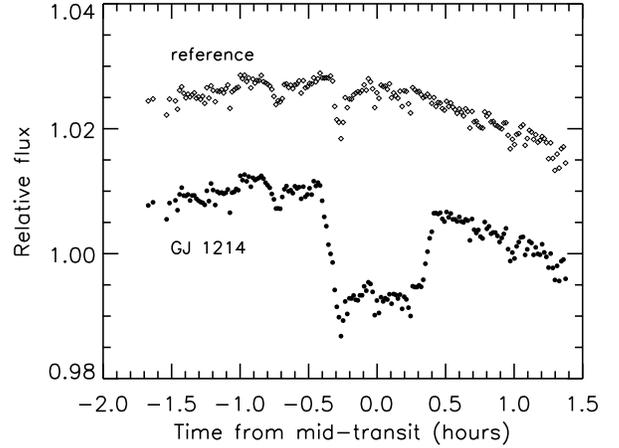}}
\caption{The broadband FORS blue raw time-series data for GJ\,1214 (filled circles) and the composite reference (open diamonds).}
\label{fig:fors_reference}
\end{figure}

This new insight to the wavelength scale of FORS grism data prompted us to re-analyze the previous data for the two transits of GJ\,1214b from \cite{bean10}, and we present the new results below. We believe that the resolution of the data can be reliably increased now that the wavelength solution has been significantly improved, and we re-bin the old spectroscopic data to 10\,nm instead of the 20\,nm used before. The reduction and correction of the data are otherwise unchanged. The basic properties of the data remain the same after the re-analysis, and the conclusions from \citet{bean10} are strengthened. In the rest of the paper we refer to the new FORS data as the FORS ``blue'' data, and the re-analyzed old data as the FORS ``red'' data.

An example spectrum of GJ\,1214 extracted from the FORS blue data is shown in Figure~\ref{fig:fors_star_spectrum}. The signal-to-noise ratio in the GJ\,1214 spectra estimated from photon counting statistics reaches upwards of 300\,pixel$^{-1}$ at 900\,nm, but only 50\,pixel$^{-1}$ at 600\,nm due to GJ\,1214's extremely red spectral energy distribution.

We created both a broadband and spectroscopic light curves from the FORS blue data by summing over the full wavelength range and over twelve channels with widths of 20\,nm, which is the lowest we feel that we can go without systematic errors given the quality of the data. We combined the data for four of the reference stars and divided the composite from the GJ\,1214 time-series to correct it. The fifth reference star was found to yield a worse correction to the photometry of GJ\,1214 and was not included in our final analysis. An illustration of the reference star correction to the broadband data is shown in Figure~\ref{fig:fors_reference}. 

As with our previous study of FORS data, we identified a slow and smooth variation in the corrected relative flux values with time superimposed on the expected transit light curve morphology. This effect is probably due to the smudges on the FORS atmospheric dispersion corrector described by \citet{moehler10}. We fit for a second-order polynomial as a function of time simultaneously with the transit modeling (see \S3) to account for this (two free parameters in addition to the normalization).

The FORS blue broadband light curve for GJ\,1214 after all corrections and decorrelations exhibits residuals from a best-fit model with a rms of 457\,ppm. This is a factor of 2.5 larger than expected from the estimated photon noise. The rms of the residuals for the spectroscopic channels range from 665\,ppm to 1785\,ppm, and the scaling from the expected levels of noise range from 1.2 to 1.4.

\begin{figure}
\resizebox{\hsize}{!}{\includegraphics{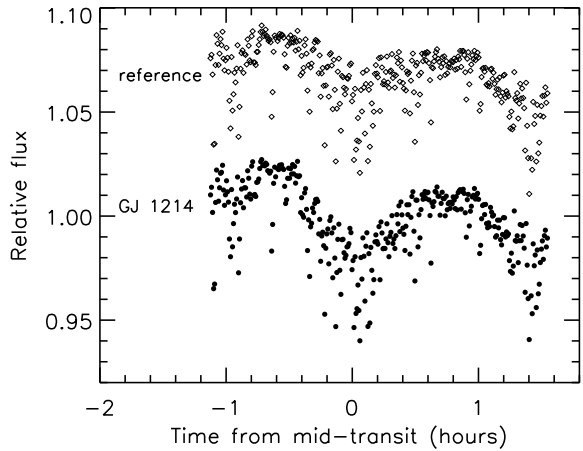}}
\caption{The HAWKI raw time-series data for GJ\,1214 (filled circles) and the composite reference (open diamonds).}
\label{fig:hawki_reference}
\end{figure}

\subsection{VLT\,+\,HAWKI}
We observed one transit of GJ\,1214b on 2011 August 10 using the standard imaging mode of the HAWKI instrument \citep{kisslerpatig08} on UT4 of the VLT facility at Paranal Observatory. The observations were done using the NB2090 filter ($\lambda_{c}$\,=\,2095\,nm, width\,=\,20\,nm). In contrast to previously reported HAWKI observations of exoplanet transits \citep[e.g., ][]{gillon09,anderson10,gibson10}, we chose to maintain a single pointing throughout the observations instead of using a dither pattern. This ``staring'' approach has been used for observations with a similar instrument to obtain the highest-precision ground-based transit light curves in the near-infrared up to now \citep[approximately 700\,ppm\,minute$^{-1}$; e.g.,][]{croll10a,croll10b,croll11b}. We reach a similar level of precision in our data using the staring approach, and this suggests that the single pointing method is indeed the best technique to use for exoplanet transit observations with HAWKI.

Observations began at UT 01:03 and continued for 2.68\,h to UT 03:44. The field was at an airmass of 1.15 at the beginning, and it set to an airmass of 1.64 by the end. The exposure times were 1.6762\,s, which is the minimum possible for the normal mode of HAWKI. The telescope was defocussed slightly to keep the counts for all the stars in the field-of-view within the linearity range of the detector. To improve the duty cycle, six separate exposures were averaged on the chip to produce a single recorded image. The total time to integrate the exposures, read out the data, and reset the detector was 28\,s. A total of 337 images were obtained, 110 of which were during transit. The maximum counts for GJ\,1214, which was the brightest star in the field-of-view, ranged from 9,000 to 25,000\,$e^{-}$. This is well below the the 1\% non-linearity level for the HAWKI detectors (approximately 60,000\,$e^{-}$) and no non-linearity corrections were applied to the data. The seeing reported by the observatory atmospheric station varied from 0.9 to 3.2\arcsec\ during the observations, but we recall that the guide probe of the telescope never registered more than 1.6\arcsec\ (this information is not included in the image headers).

After bias subtraction and flat fielding the HAWKI images, we measured aperture photometry for GJ\,1214 and 11 other stars. The photometric aperture and sky annulus sizes were optimized to give the lowest dispersion in the light curve for GJ\,1214. We settled on a stellar aperture radius of 26 pixels, and a sky annulus inner radius of 31 pixels and outer radius of 41 pixels for the final analysis.

\begin{figure*}
\resizebox{\hsize}{!}{\includegraphics{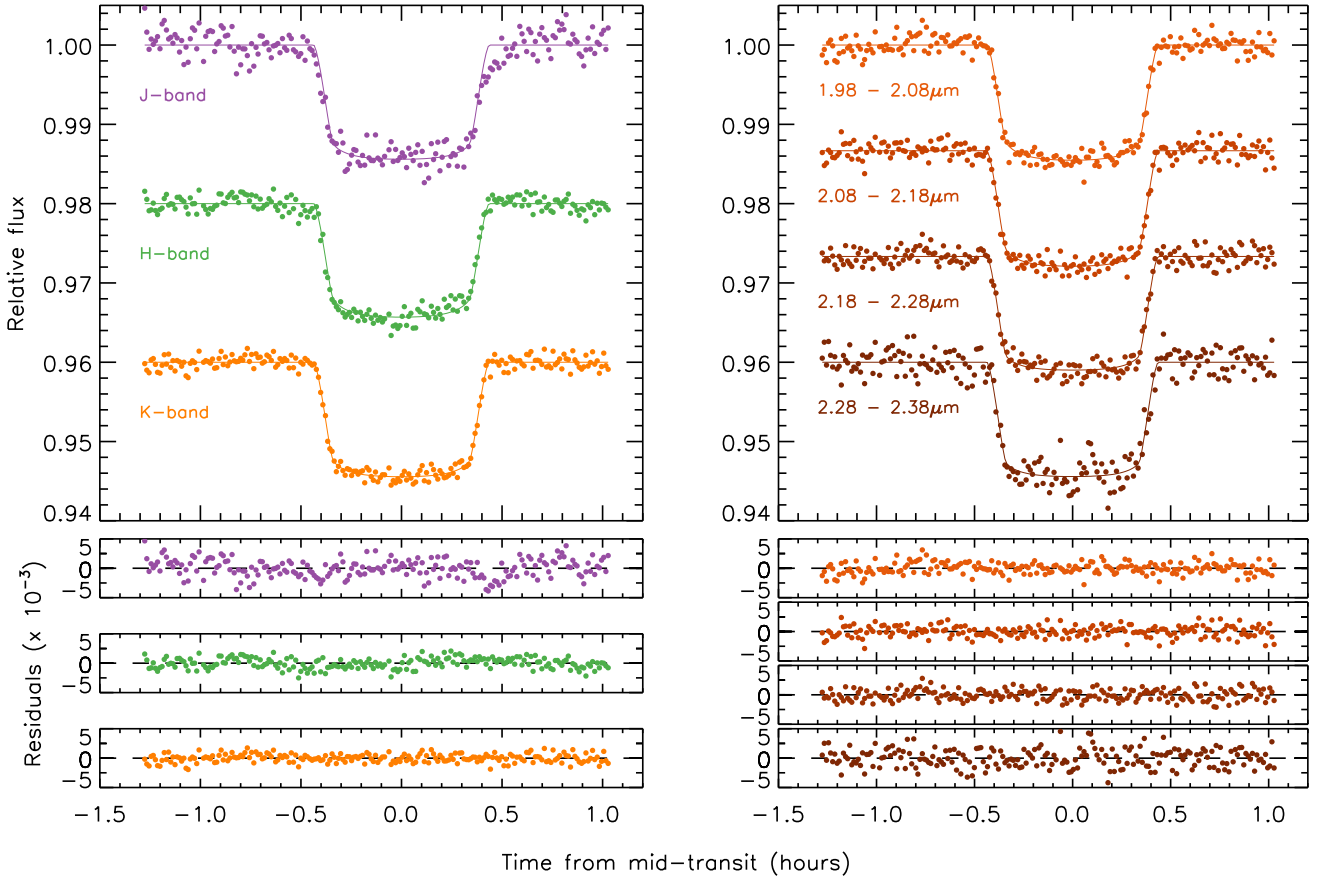}}
\caption{\textit{Upper panels} Normalized broadband (left panel) and $K$-band spectroscopic (right panel) light curves from the MMIRS data (circles). The best-fit models are shown as the solid lines. \textit{Bottom panels} Residuals from the fits (circles).}
\label{fig:mmirs_lc}
\end{figure*}

We corrected the time-series photometry of GJ\,1214 by dividing out the sum of a number of the reference stars. We experimented with different combinations of the reference stars for the correction, and we ultimately used a composite of four of them for the final analysis. A plot of the raw time-series for GJ\,1214 and the composite reference is shown in Figure~\ref{fig:hawki_reference}.

\begin{deluxetable}{llcc}
\tabletypesize{\scriptsize}
\tablecolumns{4}
\tablewidth{0pc}
\tablecaption{Mid-transit Times}
\tablehead{
 \colhead{Instrument} &
 \colhead{Epoch\tablenotemark{a}} &
 \colhead{$T_{c}$ (BJD$_{\mathrm{TDB}}$)\tablenotemark{b}} &
 \colhead{O - C (s)\tablenotemark{c}}
}
\startdata
MMIRS & 464 & 2455699.832866\,$\pm$\,6.9E-5 & -7.6\,$\pm$\,5.9 \\
FORS  & 493 & 2455745.664729\,$\pm$\,6.0E-5 & +3.1\,$\pm$\,5.2 \\
HAWKI & 517 & 2455783.59440\,$\pm$\,1.1E-4  & -0.4\,$\pm$\,9.5
\enddata
\tablenotetext{a}{Integer number of orbital periods since the transit on 2009 May 15.}
\tablenotetext{b}{BJD$_{\mathrm{TDB}}$ is the the Barycentric Julian Date in the Barycentric Dynamical Time standard \citep{eastman10}.}
\tablenotetext{c}{Residuals form the ephemeris given in \S3.}
\label{tab:times}
\end{deluxetable}

After the reference star correction, we noticed a slow and smooth variation in the corrected GJ\,1214 relative flux values with time superimposed on the expected transit light curve morphology. This trend is potentially related to the large color difference between GJ\,1214 and the reference stars, although the bandpass of the utilized filter is relatively narrow. Similar trends are regularly seen in ground-based near-infrared transit photometry \citep[e.g.,][]{croll10a,croll10b,croll11a,croll11b,sada10}, so this probably arises from a property of Earth's atmosphere. We found that the trend in our data was best modeled by a second order polynomial with time, and we fit for such a function simultaneously with the transit modeling (see \S3, two free parameters in addition to the normalization). Similar results, but with higher dispersion in the light curve residuals and larger determined parameter uncertainties, are obtained when decorrelating the data with a linear trend as a function of airmass. The data from three of the images were found to deviate by more than five times the standard deviation of the residuals from a best-fit model. These discrepant points were removed for the final analysis.

The HAWKI light curve for GJ\,1214 after all corrections, decorrelations, and trimming of discrepant points exhibits residuals from a best-fit model with a rms of 1259\,ppm. The rms of the residuals is 732\,ppm when the data are binned to 60\,s. The dispersion of the per-image residuals in the un-binned data is a factor of 1.3 larger than expected from the estimated photon noise.

\begin{figure*}
\resizebox{\hsize}{!}{\includegraphics{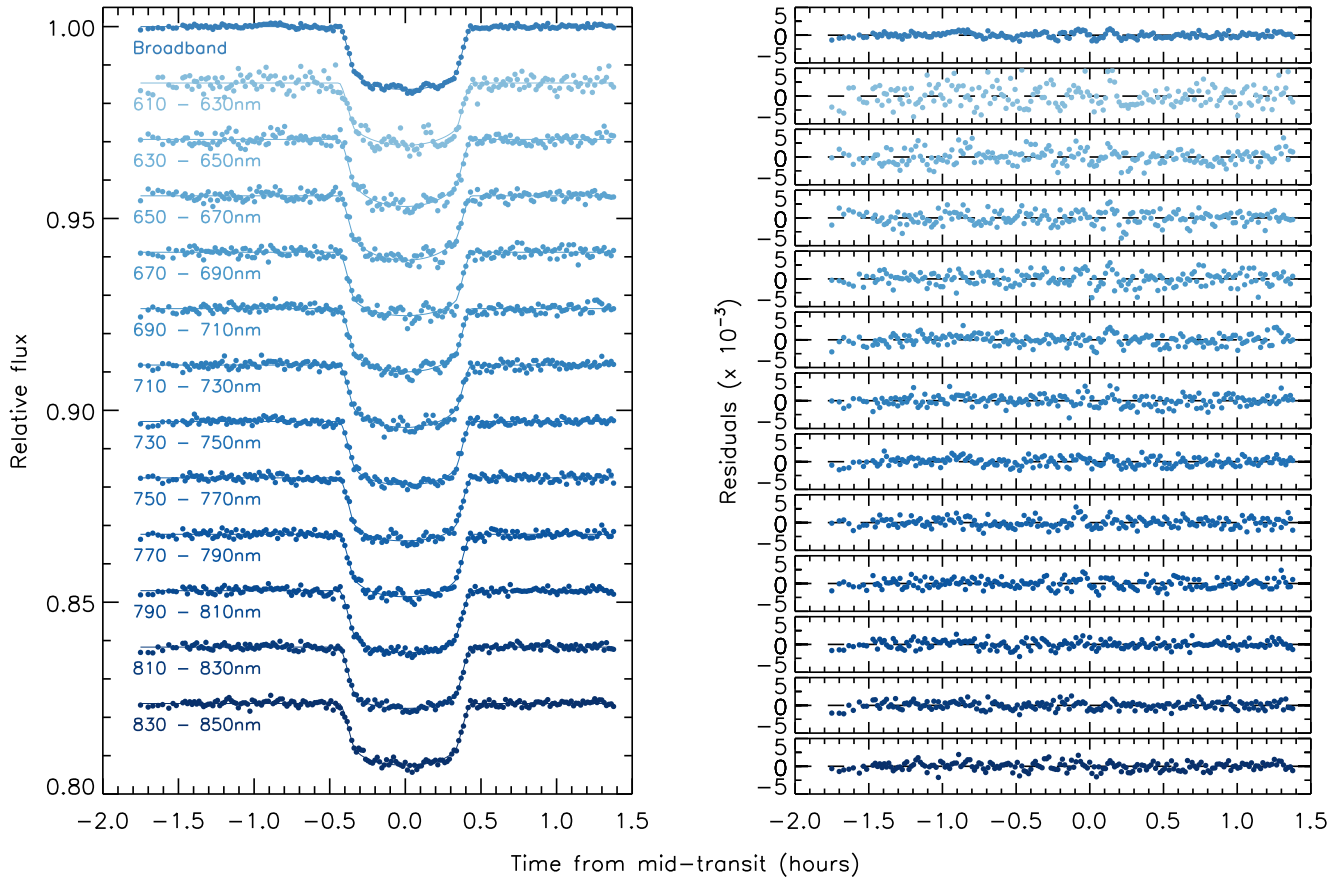}}
\caption{\textit{Left panel} Normalized light curves from the FORS blue data (circles) and best-fit models (lines). \textit{Right panel} Residuals from the fits (circles).}
\label{fig:forsb_lc}
\end{figure*}

\section{ANALYSIS}
We fit the broadband and spectroscopic light curves for GJ\,1214 with a transit model multiplied by a normalization factor and systematic decorrelation functions to investigate the transmission spectrum of its transiting planet. We used the exact analytic formulas including quadratic limb darkening given by \citet{mandel02} for the transit model. The transit model was parameterized by the planet-to-star radius ratio ($R_{p}/R_{\star}$), system scale ($a/R_{\star}$), orbital inclination of the planet ($i$), mid-transit time of the planet ($T_{c}$), and the stellar quadratic limb darkening coefficients ($\gamma_{1}$ and $\gamma_{2}$). The orbital period of the planet was fixed to the revised value given below, and the eccentricity was fixed to zero. The normalization factor for each data set was always a free parameter. The decorrelation functions described in \S2 require two additional free parameters for each wavelength bin.

We estimated limb darkening coefficients appropriate for transits of GJ\,1214b using spherically symmetric stellar model atmospheres calculated with the PHOENIX code \citep{hauschildt99}. We adopted the stellar parameters for the baseline model $T_{eff}$ = 3026\,K, [M/H] = 0.0, log\,g = 5.0, and $R_{\star}$ = 0.21\,$R_{\odot}$. We also computed models at the baseline effective temperature $\pm$\,130\,K. For the MMIRS and FORS spectroscopic data, the model intensities at each wavelength were scaled to the actual counts in the utilized spectra for GJ\,1214 to account for the unique bandpass and flux weighting of the resulting photometry. For the HAWKI photometric data, we estimated the effective observed stellar spectrum by multiplying the baseline theoretical stellar spectrum by a theoretical model for the transmission of Earth's atmosphere and by the transmission curve of the utilized filter. The theoretical model for the transmission of Earth's atmosphere was calculated for the line of sight through atmosphere above the observatory on the night of the observations using the method described by \citet{seifahrt10}. After integrating over the effective bandpasses, we fit a quadratic function to the intensities to determine the limb darkening coefficients. We limited our fits to the range $\mu\,\equiv\,cos\,\theta\,<\,0.1$, where $\theta$ is the angle between the emergent intensity and the line of sight. 

We allowed the limb darkening coefficients to be free parameters in the light curve modeling, and we used the estimated theoretical values as priors. That is, the differences between the coefficients used for the light-curve modeling and the coefficients calculated from the PHOENIX models were included in the tabulation of the goodness-of-fit metric (see below). The adopted uncertainties for the priors were set by the spread in the values for the coefficients at the three different temperatures considered for GJ\,1214. We also tested the effects of using the \citet{claret00} non-linear limb darkening law. While this functional form yielded better fits at $\mu\,<\,0.1$ for the MMIRS data, it did not yield better light curve fits and the final determined transmision spectrum values for the planet are not significantly affected by using it instead of the quadratic law. The best-fit limb darkening coefficients were in all cases within 2\,$\sigma$ of the theoretical values estimated from the $T_{eff}$ = 3026\,K model atmosphere.

We first analyzed the broadband light curves (MMIRS $J$, $H$, and $K$; FORS blue and red; and HAWKI) to determine values for all the transit parameters. We analyzed the MMIRS, FORS, and HAWKI data separately. We assumed the system scale, inclination, and transit time are the same for the $J$-, $H$-, and $K$-band MMIRS data because they were obtained simultaneously for the same transit. Each of the MMIRS bands were allowed to have a unique transit depth. We used a non-linear least-squares algorithm \citep{markwardt09} to identify the best-fit parameters, and a residual permutation bootstrap algorithm to asses the uncertainties on the parameters. The standard $\chi^{2}$ metric was used throughout to asses the quality of the model fits.

\begin{deluxetable}{lrclc}
\tabletypesize{\scriptsize}
\tablecolumns{5}
\tablewidth{0pc}
\tablecaption{Photometric Bandpasses \& Transit Depths for the FORS Blue Data}
\tablehead{
 \colhead{Band} &
 \multicolumn{3}{c}{Wavelength (nm)} &
 \colhead{$R_{p}/R_{\star}$\tablenotemark{a}}
}
\startdata
broadband & 610 &--& 850 & 0.1178\,$\pm$\,0.0007 \\
channel  1 &  610 &--& 630 & 0.1173\,$\pm$\,0.0018 \\
channel  2 &  630 &--& 650 & 0.1195\,$\pm$\,0.0012 \\
channel  3 &  650 &--& 670 & 0.1167\,$\pm$\,0.0011 \\
channel  4 &  670 &--& 690 & 0.1194\,$\pm$\,0.0011 \\
channel  5 &  690 &--& 710 & 0.1169\,$\pm$\,0.0009 \\
channel  6 &  710 &--& 730 & 0.1164\,$\pm$\,0.0009 \\
channel  7 &  730 &--& 750 & 0.1182\,$\pm$\,0.0007 \\
channel  8 &  750 &--& 770 & 0.1187\,$\pm$\,0.0008 \\
channel  9 &  770 &--& 790 & 0.1172\,$\pm$\,0.0008 \\
channel 10 &  790 &--& 810 & 0.1172\,$\pm$\,0.0007 \\
channel 11 &  810 &--& 830 & 0.1183\,$\pm$\,0.0006 \\
channel 12 &  830 &--& 850 & 0.1168\,$\pm$\,0.0007
\enddata
\tablenotetext{a}{From an analysis with the transit parameters $a/R{\star}$ and $i$ fixed to 14.97 and 88.94\degr, respectively.}
\label{tab:forsb_band}
\end{deluxetable}

The determined transit times from analysis of the broadband data with all parameters free are given in Table~\ref{tab:times}. Combining these times with the previously reported times from \citet{desert11}, \citet{carter11}, and \citet{berta11}, we can calculate a revised period $P$\,=\,1.58040481\,$\pm$\,1.2E-7\,d, and reference transit time $T_{c}$\,=\,2454966.525123\,$\pm$\,0.000032\,BJD$_{\mathrm{TDB}}$. The new transit times in this paper deviate by less than 1.3$\sigma$ from the predictions of this ephemeris, and none of other transit times exhibit deviations greater than 2.4\,$\sigma$.

The best-fit values of the system scale and the planet's orbital inclination for the broadband data are consistent with previously determined values \citep[e.g.,][]{carter11,berta11}. We fixed these parameters to the values used by \citet[][$a/R_{\star}$\,=\,14.9749, $i$\,=\,88.94$\degr$]{bean10} to measure the transit depth as a function of wavelength in the subsequent analyses. These values were also adopted by \citet{desert11} and \citet{croll11a} in their studies, so we can directly compare our results with theirs. We also fixed all the transit times to the predicted values (not best-fit) from the ephemeris given above. This is justified because there is no evidence of transit timing variations for this system.

\begin{deluxetable}{lrclc}
\tabletypesize{\scriptsize}
\tablecolumns{5}
\tablewidth{0pc}
\tablecaption{Photometric Bandpasses \& Transit Depths for the FORS Red Data}
\tablehead{
 \colhead{Band} &
 \multicolumn{3}{c}{Wavelength (nm)} &
 \colhead{$R_{p}/R_{\star}$\tablenotemark{a}}
}
\startdata
broadband & 780 &--& 1000 & 0.1168\,$\pm$\,0.0006 \\
channel  1 &  780 &--& 790 & 0.1167\,$\pm$\,0.0009 \\
channel  2 &  790 &--& 800 & 0.1160\,$\pm$\,0.0007 \\
channel  3 &  800 &--& 810 & 0.1156\,$\pm$\,0.0007 \\
channel  4 &  810 &--& 820 & 0.1176\,$\pm$\,0.0008 \\
channel  5 &  820 &--& 830 & 0.1176\,$\pm$\,0.0007 \\
channel  6 &  830 &--& 840 & 0.1162\,$\pm$\,0.0007 \\
channel  7 &  840 &--& 850 & 0.1172\,$\pm$\,0.0007 \\
channel  8 &  850 &--& 860 & 0.1151\,$\pm$\,0.0008 \\
channel  9 &  860 &--& 870 & 0.1168\,$\pm$\,0.0007 \\
channel 10 &  870 &--& 880 & 0.1171\,$\pm$\,0.0007 \\
channel 11 &  880 &--& 890 & 0.1171\,$\pm$\,0.0007 \\
channel 12 &  890 &--& 900 & 0.1159\,$\pm$\,0.0007 \\
channel 13 &  900 &--& 910 & 0.1167\,$\pm$\,0.0006 \\
channel 14 &  910 &--& 920 & 0.1175\,$\pm$\,0.0007 \\
channel 15 &  920 &--& 930 & 0.1178\,$\pm$\,0.0006 \\
channel 16 &  930 &--& 940 & 0.1165\,$\pm$\,0.0009 \\
channel 17 &  940 &--& 950 & 0.1168\,$\pm$\,0.0008 \\
channel 18 &  950 &--& 960 & 0.1176\,$\pm$\,0.0009 \\
channel 19 &  960 &--& 970 & 0.1172\,$\pm$\,0.0009 \\
channel 20 &  970 &--& 980 & 0.1161\,$\pm$\,0.0010 \\
channel 21 &  980 &--& 990 & 0.1165\,$\pm$\,0.0011 \\
channel 22 &  990 &--&1000 & 0.1168\,$\pm$\,0.0011
\enddata
\tablenotetext{a}{From an analysis with the transit parameters $a/R{\star}$ and $i$ fixed to 14.97 and 88.94\degr, respectively.}
\label{tab:forsr_band}
\end{deluxetable}

We performed two types of analyses to measure the planet's transmission spectrum. We used non-linear least-squares fitting with a residual permutation bootstrap to analyze the broadband light curves. For the spectroscopic data sets, we used a Markov Chain Monte Carlo (MCMC) algorithm to find the most likely parameter values and their uncertainties. The adopted photometric errors for the MCMC analysis were the photon-limited errors adjusted upwards to yield a reduced $\chi^{2}$\,=\,1 for the best fit. The free parameters for each wavelength bin in the two analyses were $R_{p}/R_{\star}$, quadratic limb-darkening coefficients (with priors), flux normalization, and decorrelation parameters. There were a total of six free parameters for each wavelength bin, and two ``observations'' per bin in addition to the light curves (the limb darkening priors). 

The motivation for the different analyses between the broadband and spectroscopic data sets is that they are used in different ways to investigate the planet's transmission spectrum. The broadband data are useful when compared to theoretical models in aggregate because they span a wide range of wavelengths. For this kind of study, the correlated, or red, noise in each data set must be considered because the observations were obtained for different transits with different instruments over the course of years. On the other hand, the spectroscopic data sets each tightly constrain the planet's transmission spectrum on their own. These measurements were obtained simultaneously with the same instrument for the same transit and over a limited wavelength range. In this case, the major sources of correlated noise, variations in Earth's atmospheric transparency and stellar spot crossing, are quite similar from channel to channel within a data set. Therefore, a full correlated noise analysis is not appropriate when the resulting planet transmission spectra for each spectroscopic data set are only examined in isolation, as we do in \S4.

\begin{figure}
\resizebox{\hsize}{!}{\includegraphics{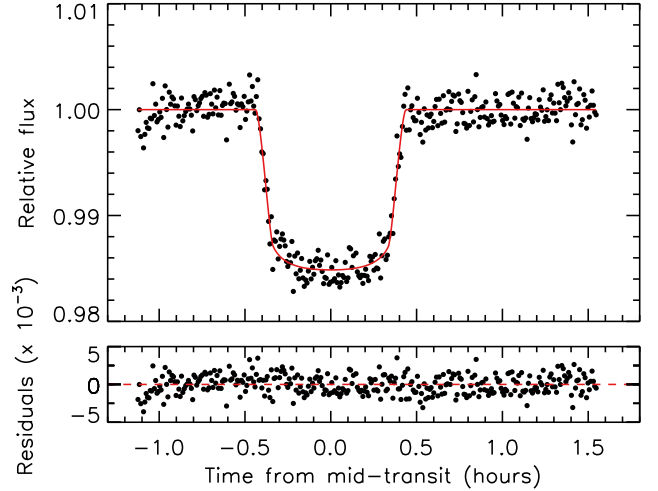}}
\caption{\textit{Upper panel} The normalized HAWKI light curve (circles) and best-fit model (line). \textit{Lower panel} Residuals from the best-fit model (circles).}
\label{fig:hawki_lc}
\end{figure}

The determined $R_{p}/R_{\star}$ values for the MMIRS, FORS blue, and FORS red data are given in Tables~\ref{tab:mmirs_band}, \ref{tab:forsb_band}, and \ref{tab:forsr_band}, respectively. The determined $R_{p}/R_{\star}$ for the HAWKI data is 0.1179\,$\pm$\,0.0012. The normalized light curves with systematics removed along with the best-fit transit models and residuals from the models are shown in Figures~\ref{fig:mmirs_lc} (MMIRS data), \ref{fig:forsb_lc} (FORS blue data), and \ref{fig:hawki_lc} (HAWKI data). The re-analyzed FORS red data do not look substantially different than those data shown by \citet[][spectroscopic data]{bean10} and \citet[][broadband data]{berta11}. 

There is some indication of the planet crossing spots on the surface of GJ\,1214 in the FORS blue data. We performed tests to investigate the influence of masking the possible spot crossing events from the light curves. The determined depths of the light curves changed depending on which points were masked, but the variation was always less than the quoted uncertainties on the $R_{p}/R_{\star}$ values, and the relative values of the FORS blue transmission spectrum were unchanged. The results presented here are from an analysis with none of the light curve points masked.

\begin{figure*}
\resizebox{\hsize}{!}{\includegraphics{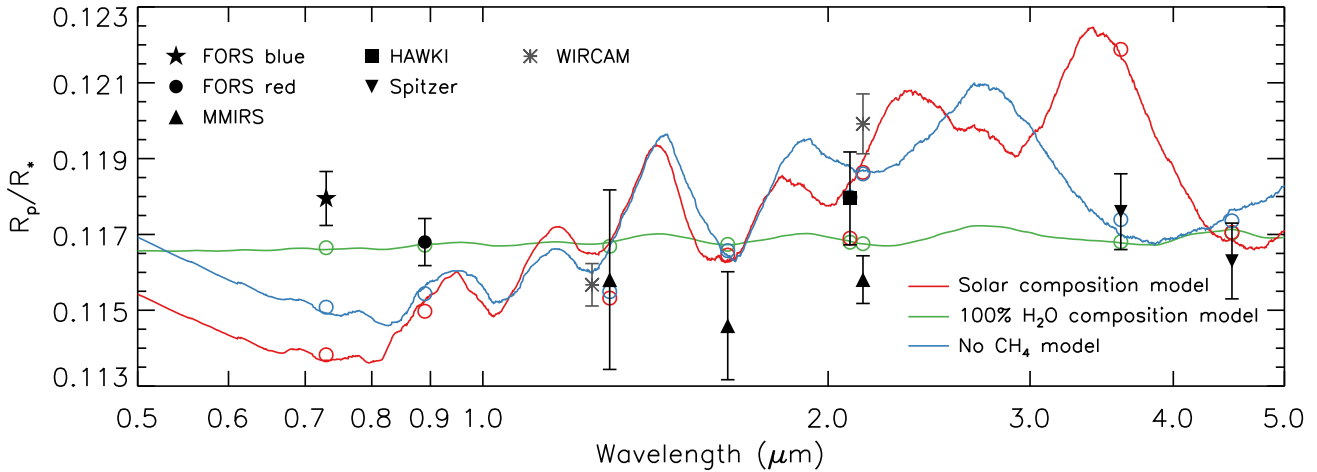}}
\caption{Broadband measurements of the transmission spectrum of GJ\,1214b (filled points) compared to theoretical models (lines). The models were binned over the bandpasses of the measurements (open circles) and scaled to give the best fit to the data. The WIRCAM $J$- and $K$-band points from \cite[][grey asterisks]{croll11a} were not included in the fitting. All calculations were done with high-resolution models; the models shown are smoothed for clarity.}
\label{fig:compare_broad}
\end{figure*}

\section{DISCUSSION}
We compare our newly measured transit depths to previous results and the predictions of theoretical models to investigate the nature of GJ\,1214b's atmosphere in this section. We utilize the models presented by \citet{millerricci10} throughout. The goals of our investigation are to determine the composition of the planet's atmosphere, and to provide an independent assessment of the transit depth in the $K$-band to test the claim from \citet{croll11a} that the transit is significantly deeper at these wavelengths. 

We look at the broadband, and the optical and $K$-band spectroscopic data sets separately. The spectroscopic data enable a search for spectral features within the different regions they cover. The strength of these data sets is that they yield extremely precise measurements on an internal scale, and each one tightly constrains the planet's transmission spectrum. However, there are offsets between the different spectroscopic data sets with sizes that are on the order of the uncertainties of the respective broadband points. The FORS blue and red data can be combined because the offset between the data sets can be determined from the values at the wavelengths where they overlap. The FORS and MMIRS spectroscopy can not be combined because the offset between the data sets is uncertain.

The broadband data cover a wide span of wavelengths, and thus probe different sources of opacity and different atmospheric pressures. These data were taken at different epochs and with different instruments, but can be combined because we considered the correlated noise in each of the data sets, and because the activity-related photometric variability of GJ\,1214 itself should only cause variations in the transit depth that are much smaller than the measurement uncertainties \citep{berta11}. The excellent agreement between the FORS blue and red observations, which were separated by a year in time, at the wavelengths the data sets have in common (see \S4.3) is additional evidence that stellar activity is not a significant problem for this investigation.

\subsection{Broadband photometry}
The broadband measurements of GJ\,1214b's transmission spectrum from \citet[][\textit{Spitzer} data]{desert11}, \citet[][ground-based near-infrared]{croll11a}, and this paper are shown compared to theoretical models for the planet's atmosphere in Figure~\ref{fig:compare_broad}. The broadband transmission spectrum from the combination of the FORS blue and red, MMIRS, HAWKI, and \textit{Spitzer} data does not exhibit any significant features. The maximum deviation from the weighted mean of the points is 1.7\,$\sigma$. 

Up until now, the only candidate detection of spectral features in GJ\,1214b's transmission spectrum has been in the $K$-band by \citet{croll11a}. Our HAWKI data point in a narrow window of the $K$-band does not have the precision to provide a strong test of the \citet{croll11a} measurement on its own. However, the broad $K$-band data point created from the combination of the MMIRS spectroscopic data has excellent precision and can provide this test. The MMIRS $K$-band point is 4.1\,$\sigma$ ($\Delta$\,$R_{p}/R_{\star}$\,=\,0.0041\,$\pm$\,0.0010) lower than the \citet{croll11a} $K$-band point, and is consistent with all of our group's measurements at other wavelengths. Therefore, we do not confirm the detection of spectral features by \citet{croll11a}.

The theoretical models for the planet's atmosphere we compare to the observations were scaled to give the best-fit to the FORS blue and red, MMIRS, HAWKI, and \textit{Spitzer} points. There are eight data points and one free parameter, and thus seven degrees of freedom for the model fits. A model for the planet's atmosphere with a 100\% water composition, which is essentially flat at the level of the precision of the measurements, yields $\chi^{2}$\,=\,10.1. On the other hand, a model for the planet's atmosphere with a solar composition (i.e., hydrogen-dominated) and assuming chemical equilibrium gives $\chi^{2}$\,=\,83.9. This suggests that the simple solar composition model is ruled out at 7.9\,$\sigma$ confidence. This strengthens the conclusions we reached previously using subsets of the current data set \citep{bean10,desert11}.

Methane is an important opacity source for solar composition models of GJ\,1214b's cool atmosphere, particularly in the $K$-band and at 3.6\,$\mu$m. However, the predicted abundance of this molecule is subject to significant uncertainty because it could be affected by non-equilibrium processes (e.g., photochemistry, thermal chemistry, and mixing). \citet{millerricci11} studied this possibility and concluded that methane depletion can at most only reach atmospheric pressure levels of 10$^{-4}$\,bar, which is far above the altitude probed by transmission spectroscopy (approximately 0.01 to 1\,bar). Similar conclusions have been reached about the difficulty of photochemistry to significantly alter the methane abundance at observable depths in the atmosphere of the warmer planet GJ\,436b \citep{line11}. Comparing a solar-composition model with methane artificially removed to our group's broadband data for GJ\,1214b gives $\chi^{2}$\,=\,43.8, which is discrepant at 5.2\,$\sigma$.

A grey broadband optical opacity source is needed for hydrogen-dominated models of the planet's atmosphere to be consistent with the data. Such a source could be high-altitude clouds or haze. The data constrain the altitude of clouds or haze in GJ\,1214b's atmosphere to be at least above a level corresponding to pressures of approximately 0.1\,bar. If methane is depleted down to a similar level, then the clouds or haze would only have to be optically thick for wavelengths less than 1\,$\mu$m.

\begin{figure}
\resizebox{\hsize}{!}{\includegraphics{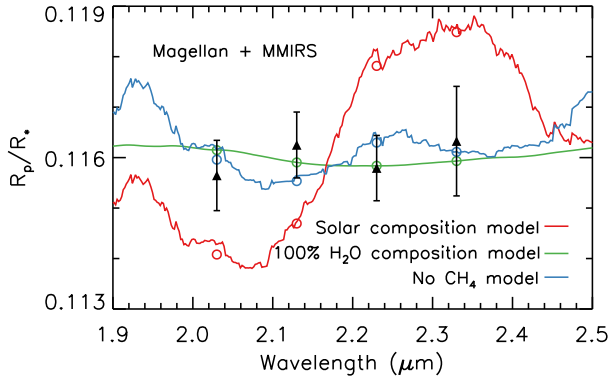}}
\caption{The derived transmission spectrum of GJ\,1214b from the MMIRS $K$-band spectroscopy (filled triangles) compared to theoretical models (lines). The models were binned over the bandpasses of the measurements (open circles) and scaled to give the best fit to the data. All calculations were done with high-resolution models; the models shown are smoothed for clarity.}
\label{fig:compare_kband}
\end{figure}

\begin{figure*}
\resizebox{\hsize}{!}{\includegraphics{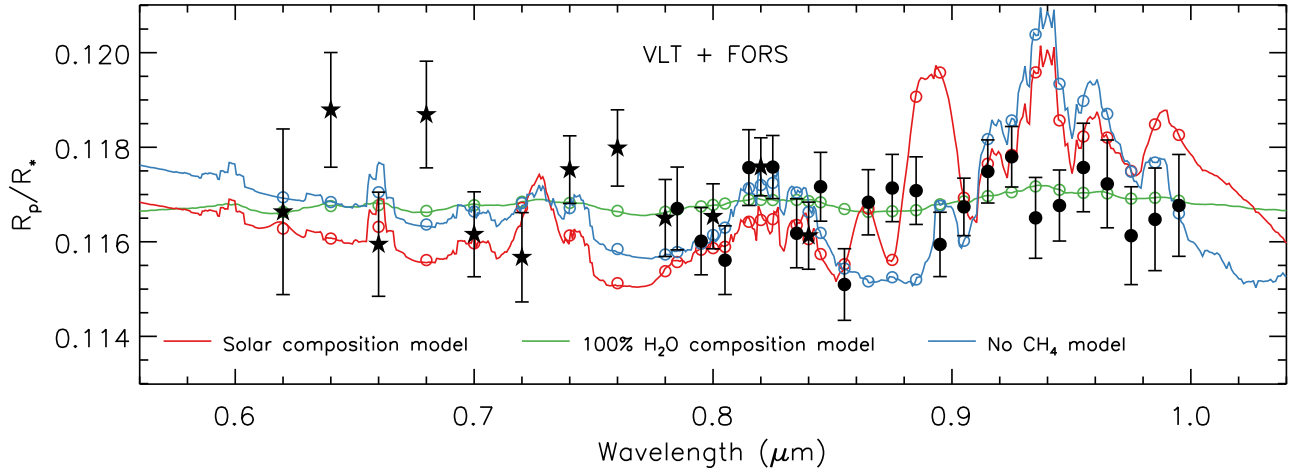}}
\caption{The derived transmission spectrum of GJ\,1214b from the FORS blue data (filled stars) and FORS red data (filled circles) spectroscopy compared to theoretical models (lines). The FORS blue were adjusted downward by 0.0007 to match the red data in the region where the data sets overlap. The models were binned over the bandpasses of the measurements (open circles) and scaled to give the best fit to the data. All calculations were done with high-resolution models; the models shown are smoothed for clarity.}
\label{fig:compare_optical}
\end{figure*}

\subsection{$K$-band spectroscopy}
In addition to providing broadband measurements, the MMIRS data also yield a low-resolution spectrum of GJ\,1214b in the $K$-band. These data are shown compared to theoretical models in Figure~\ref{fig:compare_kband}. We examine these data in isolation, and adjust the models to give the best fit without consideration for the measurements at other wavelengths. There are three degrees of freedom for the model comparisons. 

The MMIRS $K$-band spectrum is featureless and flat like the broadband data. The maximum deviation from the weighted mean of the points is 0.5\,$\sigma$. So not only do we not confirm in our broadband data the deeper $K$-band transit suggested by \citet{croll11a}, we also do not detect any spectral features in the planet's atmosphere at these wavelengths using differential spectroscopy. \citet{crossfield11} previously presented a non-detection of spectral features in this region based on high-resolution differential spectroscopy, and we confirm their results at lower resolution.

The 100\% water atmosphere model is consistent with the data ($\chi^{2}$\,=\,0.9), while the solar composition model in equilibrium is ruled out at 4.3\,$\sigma$ confidence using these data alone ($\chi^{2}$\,=\,24.5). The main opacity source in this window is methane, and the solar composition model with methane artificially removed is consistent with the data ($\chi^{2}$\,=\,2.0). The conclusion we draw from these data is that the planet's atmosphere must either be metal-enhanced or methane-depleted.

\subsection{Optical spectroscopy}
The optical spectroscopic measurements from the new FORS blue data and the re-analyzed FORS red data are shown with theoretical models in Figure~\ref{fig:compare_optical}. We examine these data in isolation, and adjust the models to give the best fit without consideration for the measurements at other wavelengths. We find that the FORS blue values are on average 0.0007 larger (indicating a deeper transit) than the FORS red data for the wavelength range the two data sets have in common. This is less than the 1\,$\sigma$ uncertainty (0.0011) between the broadband points for each data set. We subtracted this determined offset from the FORS blue data for display in Figure~\ref{fig:compare_optical} and the calculation of the fit quality for the different models. The FORS optical data include 34 measurements, and there are 33 degrees of freedom for the examination of the model fit quality.

The maximum deviation from the weighted mean of the FORS spectroscopic points is 2.3\,$\sigma$, which suggests that no spectral features are detected in these data. The solar composition model in equilibrium is ruled out at 6.6\,$\sigma$ confidence ($\chi^{2}$\,=\,115.7) using these data alone. The solar composition model with methane removed is discrepant from the FORS data at the 5.2\,$\sigma$ level ($\chi^{2}$\,=\,91.2). The 100\% water composition model is consistent with the data ($\chi^{2}$\,=\,32.4). Water mass fractions of 70\% are needed to bring the models within 3\,$\sigma$ of the data, suggesting a highly metal-rich atmosphere. This conclusion is the same as we found before using only the FORS red data at lower resolution \citep{bean10}. The higher resolution and precision for the FORS red data enabled by our improved data reduction, and the addition of the blue data tightens the constraints on the planet's optical transmission spectrum.

\section{CONCLUSIONS}
We have obtained new ground-based measurements of the transmission spectrum of the 6.5\,$M_{\oplus}$ planet GJ\,1214b in the optical between 0.61 and 0.85\,$\mu$m, and in the $J$, $H$, and $K$ near-infrared windows. We have also re-analyzed our previously reported red optical spectroscopy for the planet between 0.78 and 1.00\,$\mu$m. We were able to push to higher resolution transmission spectrum measurements with these data because of improvements in the data reduction algorithm. 

We combined the new data with previously reported measurements for GJ\,1214b's transmission spectrum that were obtained with \textit{Spitzer} \citep{desert11}. The combined data set spans the visible to the infrared (0.6 to 4.5\,$\mu$m), which makes it one of the most complete exoplanet transmission spectra obtained to date. We compared the combined data set to the previously reported measurements of \citet{croll11a} and to theoretical models for the planet's atmosphere from \citet{millerricci10}. Our main conclusion is that there is no evidence of features in the planet's transmission spectrum, and we do not confirm the detection of a significantly deeper transit in the $K$-band by \citet{croll11a}. 

Our current knowledge of GJ\,1214b's mass and radius indicates that the planet must have an atmosphere \citep{rogers10b,nettelmann11}. Our interpretation of the featureless transmission spectrum for the planet is that its atmosphere must either have at least 70\% H$_{2}$O by mass or optically thick high-altitude clouds or haze. A simple cloud- and haze-free model for the planet's atmosphere with solar composition gas in chemical equilibrium is ruled out at high confidence.

Alternatively, our knowledge of the planet's radius could be wrong. If the planet were actually significantly smaller (approximately 2\,$R_{\oplus}$, or 6.5\,$\sigma$ from the current best estimate), then it would not necessarily need to have a substantial atmosphere. The estimate of GJ\,1214b's radius depends on the assumed distance to the system, which is used to estimate the mass of GJ\,1214 itself, and this is the link in the chain with the biggest question mark above it because the most recent estimate is based on photographic plate measurements \citep{vanaltena95}. If the parallax were approximately 131\,mas (10\,$\sigma$ greater than the current estimate), the radius of the planet determined from the light curves would be consistent with the planet having no atmosphere. A new measurement of the system's trigonometric parallax using modern technology would further our knowledge of this important planet.

\acknowledgments
We thank Bryce Croll and Bj\"orn Benneke for discussions pertaining to this work. We thank Brian McCleod and Warren Brown for sharing their knowledge about the MMIRS instrument, and Brice-Olivier Demory for information about the HAWKI instrument. J.L.B. and E.K. acknowledge funding from NASA through the Sagan Fellowship Program. B.S. acknowledges supoort from NSF-MRI grant 0723073. A.S. acknowledges support from NSF grant AST-1108860. The results presented are based on observations made with ESO Telescopes at the Paranal Observatories under program 087.C-0505; and on observations made with the 6.5 m Magellan telescopes located at Las Campanas Observatory.

{\it Facilities:} \facility{VLT:Antu (FORS), VLT:Yepun (HAWKI), Magellan:Clay (MMIRS)}

\bibliographystyle{apj}
\bibliography{ms.bib}

\end{document}